\documentclass[twocolumn,prb,showpacs,preprintnumbers,amsmath,amssymb]{revtex4}

\usepackage{graphicx}
\usepackage{dcolumn}
\usepackage{bm}

\begin{document}

\title{Colossal enhancement of upper critical fields in noncentrosymmetric
heavy fermion superconductors near quantum criticality: 
CeRhSi$_3$ and CeIrSi$_3$}

\author{Y. Tada}\email{tada@scphys.kyoto-u.ac.jp}, 
\author{N. Kawakami}
\author{S. Fujimoto}

\affiliation{Department of Physics, Kyoto University, Kyoto 606-8502,
Japan}

\newcommand{\vecc}[1]{\mbox{\boldmath $#1$}}

\begin{abstract}
Strong coupling effects on the upper critical fields $H_{c2}$ along the $c$-axis in the noncentrosymmetric heavy fermion superconductors near quantum criticality CeRhSi$_3$ and CeIrSi$_3$ are examined. For sufficiently large spin-orbit
interactions due to the lack of inversion symmetry, $H_{c2}$ is mainly determined by the orbital depairing effects. From microscopic calculations taking into account the strong spin fluctuations, we show that $H_{c2}$ is extremely enhanced as the system approaches  the quantum critical point, resulting in $H_{c2}\sim 30$T even in the case with the low transition temperature $T_c(H=0)\sim 1$K,
which well explains the huge $H_{c2}$ observed in the recent experiments. 
Our results reveal intrinsic and universal properties common to strong coupling superconductivity caused by spin fluctuations near quantum criticality.
\end{abstract}


\maketitle

Quantum criticality is a phenomenon emerging in the vicinity of
a second-order phase transition at sufficiently low temperatures
governed not by thermal fluctuations 
but by quantum fluctuations.
There has been accumulated evidence that quantum criticality 
plays an essential role in unconventional superconductivity realized in 
strongly correlated electron systems such as heavy fermion 
compounds and high-$T_c$ cuprates\cite{pap:qcp}.
In particular, it has been proposed by several authors that magnetic 
critical fluctuations mediate Cooper pairing in some classes of unconventional 
superconductors\cite{pap:Miyake,pap:Monthoux_Pines,
pap:Kirkpatrick,pap:Wang, pap:Roussev,pap:Chubukov}.
In this scenario, it is expected that 
the pairing interaction is explosively enhanced 
just at the quantum 
critical point (QCP),
though such enhancement has not been observed so far in real systems, 
partly because the QCP at zero temperature associated with
the magnetic order is veiled by the superconducting phase.
In this respect, the recent experimental studies on the upper 
critical fields of noncentrosymmetric heavy fermion superconductors 
CeRhSi$_3$\cite{pap:Kimura,pap:Muro} and 
CeIrSi$_3$\cite{pap:Sugitani,pap:Okuda} 
under applied pressure
are very curious\cite{pap:Kimura_hc2,pap:Settai}. 
In these systems, the distance from the QCP is controlled by the applied 
pressure.
The remarkable features of the experimental results are as follows: 
(i) As the pressure approaches a critical value, these systems 
exhibit extremely high upper critical fields $H_{c2}$ which 
exceed the orbital limit as well as the Pauli limit estimated by the 
conventional BCS theory. The observed $H_{c2}\sim 30$T is, surprisingly, almost
comparable to those of high-$T_c$ systems, though
the transition temperature at zero field is merely $\sim 1$K.
(ii) The increase in the upper critical field is enormously 
accelerated as the temperature $T$ is decreased, making a sharp 
contrast to any other superconductors in which the increase of 
$H_{c2}$ becomes slower 
as $T$ is decreased.
(iii) $H_{c2}$ increases very rapidly as the pressure approaches 
the critical value, though, by contrast, the pressure dependence 
of $T_c$ is moderate. 

The absence of the Pauli depairing effect in these systems is consistent
with the previous theoretical predictions\cite{pap:Frigeri}.
However, the origins of the strong suppression of the orbital 
depairing effect and the other extraordinary behaviors (ii) and (iii) 
 have not been elucidated so far.
It is particularly important to clarify how these unusual 
properties are related to quantum criticality 
and also what kind of role is played by the noncentrosymmetry in 
the compounds CeRh(Ir)Si$_3$.
 The answers to the questions should certainly provide us with deep 
understanding of superconductivity for strongly 
correlated electrons near quantum criticality.


In this letter, we demonstrate that the above striking features are 
naturally understood in terms of the strongly enhanced pairing 
interaction due to spin fluctuations  
in the vicinity of the magnetic QCP.
Our results reveal that the above features experimentally observed for  
CeRhSi$_3$ and CeIrSi$_3$ are not
specific to these noncentrosymmetric superconductors, but rather
 intrinsic and universal in strong coupling
superconductivity caused by critical fluctuations, as long as 
the Pauli depairing effect is suppressed by some other mechanisms
such as the lack of inversion center of the crystal structure as 
in the case of CeRh(Ir)Si$_3$.  Therefore, the unusual 
properties found for the upper critical fields 
uniquely characterize the interplay between quantum criticality and 
spin-fluctuation-mediated Cooper pairings.

We focus on the low energy quasiparticles in the clean limit 
mainly formed by $f$-electrons 
and approximate the system by
the following effective single-band tight-binding model,
\begin{eqnarray*}
S&=&S_0+S_{\rm SF},\\
S_0&=&
\sum_kc^{\dagger}_k\left(-i\omega_n+\varepsilon(\vecc{k})\right)c_k
+\sum_kc^{\dagger}_k
\alpha \vecc{\mathcal L}_0^H(\vecc{k})\cdot \vecc{\sigma}c_k,\\
S_{\rm SF}&=&-\sum_{kk^{\prime}q} \frac{g^2}{6}\chi(q)
\vecc{\sigma}_{\alpha \alpha^{\prime}}
\cdot \vecc{\sigma}_{\beta \beta^{\prime}}
c^{\dagger}_{k+q\alpha}c_{k\alpha^{\prime}}
c^{\dagger}_{k^{\prime}-q\beta}c_{k^{\prime}\beta^{\prime}},
\end{eqnarray*}
where $c_k=(c_{k\uparrow},c_{k\downarrow})^t$ is the annihilation
operator of the Kramers doublet.
Here we have introduced the notation $k=(i\omega_n,\vecc{k})$.
The lack of inversion symmetry leads to the spin-orbit interaction
which is represented in the second term with the Zeeman splitting.
The third term is 
the electron-electron interactions through spin fluctuations.
For CeRhSi$_3$ and CeIrSi$_3$ which have body centered 
tetragonal lattice structures,
the dispersion relation and the spin-orbit interaction are given by
$\varepsilon(\vecc{k})=-2t_1(\cos k_x+\cos k_y)+4t_2\cos k_x\cos k_y
-8t_3\cos(k_x/2) \cos(k_y/2) \cos k_z-\mu$ and 
$\alpha \vecc{\mathcal L}_0^H=(\alpha \sin k_y,-\alpha \sin k_x,-\mu_B H)$.
We fix the parameters as
$(t_1,t_2,t_3,n,\alpha)=(1.0,0.475,0.3,1.05,0.5)$ by taking $t_1$ as
the energy unit.
The Fermi surface determined by these parameters is 
in qualitative agreement with the band calculation 
and 
can reproduce the peak structures of the momentum-dependent 
susceptibility 
observed by the neutron scattering experiments 
\cite{pap:Tada,com:Harima,pap:Aso}.
To describe the strong spin fluctuations near the QCP,
we phenomenologically introduce the effective interaction
between quasiparticles\cite{pap:Monthoux_Pines,
pap:Moriya,pap:MMP,pap:Monthoux},
\begin{eqnarray*}
\chi(i\nu_n,\vecc{q})&=&\sum_{a}
\frac{\chi_0\xi^2}{1+\xi^2(\vecc{q}-\vecc{Q}_a)^2
+|\nu_n|/(\Gamma_0\xi^{-2})}
\end{eqnarray*}
and
$
\xi(T)=\frac{\tilde{\xi}}{\sqrt{T+\theta}},
$
where $\chi_0$ and $\Gamma_0$ are respectively the susceptibility 
and the energy scale of spin fluctuations without strong correlations.
These quantities are renormalized through the coherence length $\xi(T)$ as 
the system approaches the QCP. 
The critical exponent of $\xi$ is the mean field value $1/2$
and $\theta$ is
considered to decrease monotonically as the applied pressure
approaches the critical value for the AF 
order\cite{pap:Moriya,pap:MMP}.
The temperature dependence of $\xi$ 
is also consistent with the recent NMR experiment
for CeIrSi$_3$\cite{pap:Mukuda}.
The propagating vectors are $\vecc{Q}_1=(\pm 0.43\pi,0,0.5\pi),
\vecc{Q}_2=(0,\pm 0.43\pi,0.5\pi)$ 
according to the neutron scattering experiments\cite{pap:Aso}.
In this study, we fix the parameters in $\chi(i\nu_n,\vecc{q})$ 
as $\Gamma_0=3.6$
and $\tilde{\xi}=0.4647$. The former 
is of the same order as
the Fermi energy and the latter
is determined so that the maximum of $\xi$ would be $\xi_{\rm max}\sim 
\delta k^{-1}$, where $\delta k$ is the interstice of the $\vecc{k}$-mesh
of the Brillouin zone used in our numerical calculations.

To describe the quasiparticles in the strong coupling regime,
we introduce the normal selfenergy up to the first order
in $g^2\chi_0$ which is given by
$\Sigma_{s_1s_2}(k)=\frac{T}{N}\sum_{k^{\prime}}g^2\chi(k-k^{\prime})
G_{s_1s_2}^{0}(k^{\prime})\delta_{s_1s_2},
$
neglecting off diagonal elements which are not
important in the present study.
$G^{0}$ is the Green's function for noninteracting quasiparticles
with spin indices $s_1,s_2$.
${\rm Re}\Sigma$ only gives the deformation of the Fermi surface
while ${\rm Im}\Sigma$ gives two 
important effects to the quasiparticles around the 
Fermi level; 
the mass enhancement by the factor
$z^{-1}=1-({\rm Im}\Sigma(\pi T)
-{\rm Im}\Sigma(-\pi T))/2\pi T$
and the damping factor $\gamma=-{\rm Im}\Sigma$. 
It is reasonable to consider that $\varepsilon(\vecc{k})$
already includes the shift due to ${\rm Re}\Sigma$ and to
replace $\varepsilon(\vecc{k})
+{\rm Re}\Sigma(k) \rightarrow \varepsilon(\vecc{k})$.
We also neglect the change in $\vecc{\mathcal L}_0^H$ by
the normal selfenergy, which has little influence
in the discussion of the upper critical fields.
We note that the coupling constant $g$ should be regarded
as an effective one renormalized
by the vertex corrections\cite{pap:Yonemitsu,pap:Monthoux_ver}.

To study
the transition temperatures under applied fields,
we use the linearlized Eliashberg equations where 
the Green's functions of the quasiparticles depend on the vector potential 
$\vecc{A}$.
We employ the familiar quasiclassical approximation for
the Green's function 
$G(i\omega_n,\vecc{x},\vecc{y};\vecc{A})
=e^{ie\varphi(\vecc{x},\vecc{y})}G(i\omega_n,\vecc{x}-\vecc{y}
;\vecc{A}=0)$, $\varphi(\vecc{x},\vecc{y})=\int_{\vecc{y}}^{\vecc{x}}
\vecc{A}(\vecc{s})d\vecc{s}$
which
is legitimate for $k_Fl_H\gg 1$ where $k_F$ is the Fermi
wave number and $l_H$ the magnetic length.
We take only the $N=0$ Landau level 
because, as shown below, $H_{c2}$ is mainly determined by the orbital depairing
effect.
The gap function is approximated by 
$\Delta_{\alpha \alpha^{\prime}}(k;\vecc{R})=
\Delta_{\alpha \alpha^{\prime}}(k)\phi_0(R_x,R_y)$,
where $\phi_0$ is the lowest Landau level wave function.
The resulting Eliashberg equations for $\Delta(k)$ with
applied field $\vecc{H}=(0,0,H)$
are,
\begin{eqnarray}
\lefteqn{
\Delta_{\alpha \alpha^{\prime}}(k)
=\frac{T}{N}\sum_{k^{\prime},\beta \beta^{\prime} \gamma
\gamma^{\prime}}V_{\alpha\alpha^{\prime},\beta\beta^{\prime}}
(k,k^{\prime})} \nonumber \\
&&\sum_{\tau=\pm}\left(
\frac{1+\tau \vecc{\hat{\mathcal L}}_0^H(\vecc{k})
\cdot \vecc{\sigma}}{2}\right)_{\beta \gamma}
i{\rm sgn}(\tilde{\omega}^{\prime}) \nonumber \\
&&\left( \frac{2}{a_{\tau}(\vecc{k}^{\prime})}\right)^{1/2}
f\left( \frac{b_{\tau}(k^{\prime})}{\sqrt{2a_{\tau}(\vecc{k}^{\prime})}}\right)
G_{\beta^{\prime}
\gamma^{\prime}}(-k^{\prime})
\Delta_{\gamma \gamma^{\prime}}(k^{\prime}),
\label{eq:Eliashberg_LLL}
\end{eqnarray}
where 
$
\vecc{\hat{\mathcal L}}_0^H(\vecc{k})=\vecc{\mathcal L}_0^H(\vecc{k})/
\sqrt{\sum_{i=1}^3{\mathcal L}_{0i}^H(\vecc{k})^2}
$,
$\tilde{\omega}(k)=\omega_n-{\rm Im}\Sigma(k)$,
$a_{\tau}(\vecc{k})=\sqrt{|e|H}(v_{\tau x}(\vecc{k})^2
+v_{\tau y}(\vecc{k})^2)$,
$b_{\tau}(k)=|\tilde{\omega(k)}|
+i{\rm sgn}(\tilde{\omega})\varepsilon_{\tau}(\vecc{k})$, 
with the dispersion
$\varepsilon_{\tau}(\vecc{k})=\varepsilon(\vecc{k})+\tau \alpha
\sqrt{\sum_{i=1}^3{\mathcal L}_{0i}^H(\vecc{k})^2}$
and the velocity 
$\vecc{v}_{\tau}(\vecc{k})=\nabla \varepsilon_{\tau}(\vecc{k})$.
$f(z)$ is defined 
as
$f(z)=\frac{\sqrt{\pi}}{2}e^{z^2}{\rm erfc}(z)$.
Here, we have neglected the $\Pi$ operators in 
$\gamma=-{\rm Im}\Sigma$
because $\nabla \gamma \cdot \Pi$ represents the contribution
to $H_{c2}$ from the anisotropy of the lifetime of 
the quasiparticles on the Fermi surface and is of secondary
importance compared to $\gamma$ itself.
The velocity can be renormalized by the factor $z(\vecc{k})$, 
leading to enhancement of $H_{c2}$.
The effective pairing interaction $V$ is evaluated 
at the lowest order in $g^2\chi_0$,
\begin{eqnarray*}
V_{ss,ss}(k,k^{\prime})&=&-\frac{1}{6}g^2
       \chi(k-k^{\prime})+\frac{1}{6}g^2\chi(k+k^{\prime}),\\
V_{s\bar{s},s\bar{s}}(k,k^{\prime})&=&
      \frac{1}{6}g^2\chi(k-k^{\prime})+\frac{1}{3}g^2\chi(k+k^{\prime}),
\end{eqnarray*}
$
V_{s\bar{s},\bar{s}s}(k,k^{\prime})=-V_{s\bar{s},s\bar{s}}(k,k^{\prime})
$
and the other components 
are zero.
The applied fields will not affect $V$, because $\Gamma_0=3.6$ is
large enough compared to the energy scale of the applied fields.

\begin{figure}
\begin{center}
\includegraphics[width=2.5in]{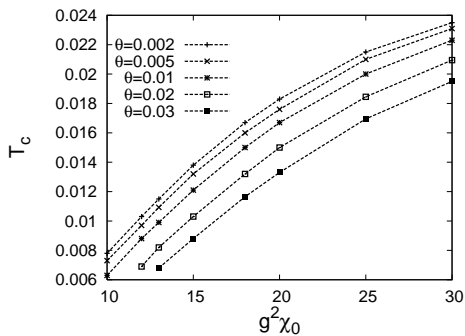}
\end{center}
\caption{\label{fig:U-T}
Transition temperatures for the extended $s$-wave state
as functions of $g^2\chi_0$
for several $\theta$.}
\end{figure}
By solving eq.(\ref{eq:Eliashberg_LLL}) at $H=0$, 
we find that the A$_1$ symmetric superconducting
state is most stable among the five irreducible representations 
of point group C$_{4v}$, which is consistent with the 
previous study\cite{pap:Tada}.
In the following, we focus on this state.
The A$_1$ symmetric order parameter is in the form of
$\Delta(i\omega_n,\vecc{k})
=[\Delta_s(i\omega_n)d_0(\vecc{k})+\Delta_t(i\omega_n)\vecc{d}(\vecc{k})
\cdot \vecc{\sigma}]
i\sigma_2$ with $d_0=\cos (2k_z),\vecc{d}=\cos (2k_z)\vecc{\mathcal L}_0$.
In our model, however, the amplitude of the triplet part $\Delta_t$
is so small ($\lesssim 0.01\Delta_s$) that we will neglect it in the following.
In Fig.\ref{fig:U-T}, we show the transition temperatures $T_c$
for the A$_1$ symmetric superconducting state at $H=0$
as functions of $g^2\chi_0$ for several $\theta$.
The transition temperatures saturate for large $g^2\chi_0$ because
of the strong coupling effects.
Their dependence on $\theta$ is quite weak.
\begin{figure}
\begin{center}
\includegraphics[width=2.5in]{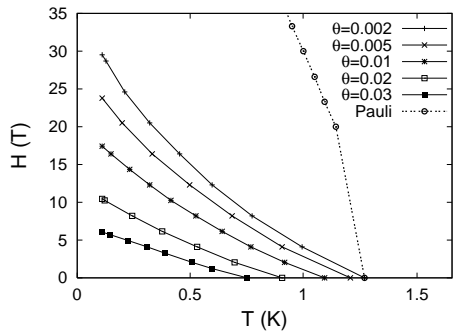}
\end{center}
\caption{\label{fig:T-H}Temperature dependence of the upper
critical fields for the extended $s$-wave state.
The dotted line with open circles 
represents the Pauli limiting field for $\theta=0.002$
and the others the orbital limiting fields for several $\theta$.
The unit of $\theta$ is $t_1=113$K.}
\end{figure}

We now turn to the case with $H\neq 0$.
In Fig.\ref{fig:T-H}, 
we
show upper critical fields as functions of temperature for 
several $\theta$ at $g^2\chi_0=13$.
For this value of $g^2\chi_0$, 
the renormalization factor $z^{-1}|_{T=0.01}$ is 
$\sim 2$, which indicates that the system is in the strong coupling regime. 
In Fig.2, we use the following values of the parameters; 
(lattice constant)=4\r A\cite{pap:Muro,pap:Okuda}
and $t_1=113$K.
The latter is obtained by identifying the calculated maximum
transition temperature $T_c=0.0115$ as 1.3K
which is the averaged value
of maximum $T_c$ for CeRhSi$_3$\cite{pap:Kimura} 
and CeIrSi$_3$\cite{pap:Sugitani}.
The dotted line with open circles in Fig.2
corresponds to the Pauli limiting field for 
$\theta/t_1=0.002$ and the other lines to
the orbital limiting fields for several $\theta$. 
In the present parameters, transition temperatures are typically 
$T_c\simeq 0.01t_1$ and thus the spin-orbit coupling constant 
is very large compared to $T_c$ ($\alpha=0.5t_c\simeq 50T_c$), leading to
$H_{c2}^{\rm Pauli}$ largely exceeding 
$H_{c2}^{\rm orb}$.
Note that, with sufficiently large $\alpha$, $H_{c2}^{\rm Pauli}(0)$
calculated from the Eliashberg equations coincides with the one
obtained from thermodynamic considerations\cite{pap:Fujimoto}.
From the result, we can
conclude that, for sufficiently large 
$\alpha \gg T_c$, the upper critical fields
are determined by the orbital limiting fields
in accordance with the previous studies\cite{pap:Frigeri,pap:Fujimoto}.
A remarkable point in Fig.\ref{fig:T-H} is that the orbital limiting fields 
are enhanced as $\theta$ is decreased, and reach 30T for sufficiently small
$\theta$ even when $T_c(H=0)\sim 1$K, which
well explains the experimentally observed huge $H_{c2}$ for
CeRhSi$_3$ and CeIrSi$_3$.
Also, the $H_{c2}$ curve 
has a upward curvature,
which is consistent with
the experimental observations.
These results verify that 
the superconductivity in 
these systems
is mediated by the strong antiferromagnetic spin fluctuations
near the QCP.
\begin{figure}
\begin{center}
\includegraphics[width=2.5in]{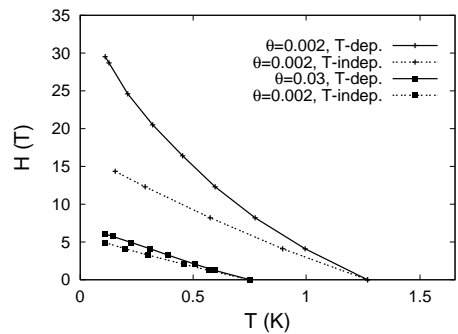}
\end{center}
\caption{\label{fig:T-H_2}
$H_{c2}^{\rm orb}$ versus $T$ calculated with $T$-dependent $\xi$(solid)
and $T$-independent $\xi$(dotted) for $\theta=0.002, 0.03$.
}
\end{figure}

The origin of the colossal enhancement in $H_{c2}^{\rm orb}$ is 
understood as follows.
As the temperature approaches absolute zero, 
the antiferromagnetic correlation length 
$\xi(T)$ increases, leading to the enhancement of the effective pairing 
interaction $V\propto \xi(T)^2$, while 
the pair breaking effect due to 
the quasiparticle damping
${\rm Im}\Sigma(T)$ is suppressed.
As a result, the coherence length of the superconducting state is
extremely reduced at low temperatures.
This behavior is significant especially for small $\theta$, 
giving rise to the huge values of $H_{c2}^{\rm orb}$, as seen in Fig.\ref{fig:T-H}. To demonstrate the importance of the rapid increase in $\xi(T)$,
we also calculate $H_{c2}$ by assuming that $\xi(T)$ is independent of temperature and has a constant value, $\xi=\xi(T_c)$ (Fig. \ref{fig:T-H_2}).
We see that the $H_{c2}^{\rm orb}$ curves calculated with
temperature independent $\xi$ exhibit neither significant upturn 
 nor enhanced values of $H_{c2}(T)$ at
low temperatures. The comparison of these results with those 
computed with the correct $\xi(T)$, as shown in Fig. \ref{fig:T-H_2}, 
concludes that the rapid increase in $\xi(T)$ is 
essential to obtain the huge values of $H_{c2}^{\rm orb}$ as well 
as the characteristic upturn profile at low temperatures.

\begin{figure}
\begin{center}
\includegraphics[width=2.5in]{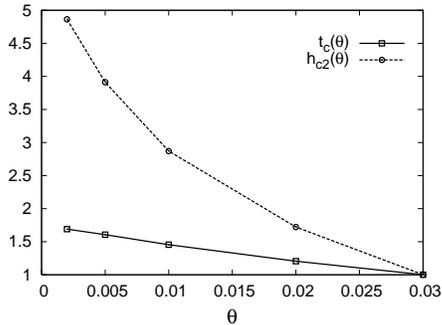}
\end{center}
\caption{\label{fig:th-T-H}
$\theta$ dependence of the transition temperature $t_c(\theta)$
and the upper critical field $h_{c2}(\theta)$.
The definitions of $t_c(\theta)$ and $h_{c2}(\theta)$
are given in the text.
}
\end{figure}
Another remarkable feature of the experimental observations
for CeRhSi$_3$ and CeIrSi$_3$ is the strong pressure dependence 
of $H_{c2}$\cite{pap:Kimura_hc2,pap:Settai}; i.e.
$H_{c2}$ increases explosively as the pressure approaches 
the critical value, while the pressure dependence 
of $T_c$ is moderate. 
To discuss this point in detail, 
we focus on $T_c(H=0)$ and $H_{c2}$ at 
low temperature as functions of $\theta$.
Figure \ref{fig:th-T-H} shows the $\theta$ dependence of
the transition temperatures at $H=0$ and the
upper critical fields at $T=T_{\rm min}\equiv 0.001$ 
normalized by the values calculated with 
$\theta=\theta_{\rm max}\equiv 0.03$;
$t_c(\theta)\equiv T_{c}(H=0,\theta)/T_c(H=0,\theta=\theta_{\rm max})$
and $h_{c2}(\theta)\equiv H_{c2}(T=T_{\rm min},\theta)/
H_{c2}(T=T_{\rm min},\theta=\theta_{\rm max})$.
We see that $t_c(\theta)$ is slightly changed as 
$\theta$ is decreased. 
By contrast, $h_{c2}(\theta)$ is strongly enhanced,
which is consistent with the above-mentioned
experimental results\cite{pap:Kimura_hc2,pap:Settai}.
These behaviors are understood as a result of
the strongly enhanced pairing interaction at low temperatures
in the vicinity of the QCP.
Although the increase of $T_c(H=0)$ near the QCP (i.e. $\theta=0$)
is considerably suppressed by the pair breaking effects due to
the inelastic scatterings with spin fluctuations,
$H_{c2}$ at low temperatures is not seriously affected by them,
resulting in the strong enhancement of $H_{c2}$ for $\theta \sim 0$.

We have theoretically clarified  the origins of several puzzling phenomena
posed experimentally for CeRh(Ir)Si$_{3}$.
Here we wish to stress that the extraordinary properties 
of $H_{c2}$ discussed here are not specific to these compounds, but are
inherent in the orbital limited superconductors caused by
spin fluctuations in the vicinity of the QCP.
However, such anomalous behaviors of $H_{c2}$
have not been observed so far in any
other systems than CeRh(Ir)Si$_3$,
although there are many superconducting 
systems which are believed to be located in the
vicinity of a magnetic QCP.
The key to resolve this apparent contradiction 
is the lack of inversion symmetry.
In the presence of the sufficiently large anisotropic spin-orbit
interaction, as discussed above, 
the dominant mechanism which determines the
properties of $H_{c2}$ is the orbital depairing.
In addition, generally, the orbital limiting fields are
much more influenced by the electron correlations than
the Pauli ones. In ordinary centrosymmetric superconductors,
the enhanced strong coupling effects on $H_{c2}$ near a QCP, which could 
exist in principle, are masked by the Pauli depairing effects. 
The noncentrosymmetric heavy fermion superconductors
CeRhSi$_3$ and CeIrSi$_3$ with  large anisotropic spin-orbit interactions
are the first systems in which $H_{c2}$ directly exhibits 
the enhanced strong coupling
effects due to spin fluctuations near the QCP. 
We thus naturally expect that the anomalous properties of $H_{c2}$
could be more generally observed in other strong-coupling
superconductors caused by critical spin fluctuations if 
the Pauli depairing effect could be suppressed by some
other mechanisms.

In summary,
we have studied the strong coupling effects on the upper critical 
fields in CeRhSi$_3$ and CeIrSi$_3$.
We have found that 
the upper critical fields 
are extremely 
enhanced by critical spin fluctuations near the QCP,
which successfully explains the recent experimental observations.
We have elucidated that the huge values of $H_{c2}$ and its strong 
dependence on the applied pressure is a universal property inherent
in the orbital limited superconductors near the QCP.

We thank N. Kimura, R. Settai and Y.\=Onuki for valuable discussions.
Numerical calculations were partially performed at the 
Yukawa institute. 
This work is partly supported by the Grant-in-Aids for
Scientific Research from MEXT of Japan
(Grant No.18540347, Grant No.19014009, Grant No.19014013, Grant No.19052003, 
and Grant No.20029013).
Y. Tada is supported by JSPS Research Fellowships for Young
Scientists.


\end{document}